\documentclass{ws-p8-50x6-00}

\def\APA{{\tt APACIC++} }
\def\AME{{\tt AMEGIC++} }
\def\AA{{\tt APACIC++/AMEGIC++} }

\begin{document}

\title{Parton Showers and Multijet Events
	\footnote{Dedicated to Prof. F. Sellschop on the occasion of
	his ${\rm 70^{th}}$ birthday.}}

\author{R. Kuhn, A. Sch{\"a}licke, F. Krauss and G. Soff}

\address{Institut f\"ur Theoretische Physik, TU Dresden, 
             01062 Dresden, Germany\\
	Max Planck Institut f{\"u}r Physik komplexer Systeme,
        	      01187  Dresden, Germany\\
	E-mail: kuhn@theory.phy.tu-dresden.de}


\maketitle

\abstracts{
A Monte-Carlo event--generator has been developed which is dedicated
to simulate electron--positron annihilations. Especially a new
approach for the combination of 
matrix elements and parton showers ensures the independence of the
hadronization parameters from the CMS energy. This enables for the
first time the description of multijet--topologies, e.g. four jet
angles, over a wide range of energy, without changing any parameter of
the model. Covering all processes of the standard model our simulator
is capable to describe experiments at present and future
accelerators, i.e. the LEP collider and a possible Next Linear Collider(NLC).}

\section{Introduction}

Multijet events play a crucial role in present and
future high energy particle physics. Already in past experiments
multijet observables have lead to proofs of the theory of
strong interaction, e.g. the underlying symmetry group has been
established. With rising energies the production of multijet events
via the electroweak interaction becomes more important, e.g. the
creation of $ZH$ and $W$-pairs involves at least four jets and
dominates the QCD background. In addition, the majority of signals for
new physics, e.g. supersymmetry is related to multijet topologies.  
        
Our Monte-Carlo generator \APA(A PArton Cascade In C++)\cite{APA} in combination
with our matrix element generator \AME(A Matrix Element Generator in
C++)\cite{AME} was developed with the aim to describe these multijet events in a
correct manner over a wide range of energy. This was achieved with a
new approach for combining 
the advantages of matrix elements and parton showers, which leads to a
good describtion of experiments at present accelerators,
i.e. the LEP collider. Including extensions of
the standard model, primarily supersymmetry, \AA will be dedicated
for the search of new physics at a possible Linear collider.

The paper is outlined as follows. Tracking the physics features
related to event generation in the subsequent sections we describe
briefly the treatment of initial state radiation, matrix element
generation and evaluation, the combination with the parton shower and
the parton shower itself.

%\newpage

\section{Initial state radiation}

\begin{figure}[h]
\begin{center}
\includegraphics[width=7cm,angle=270]{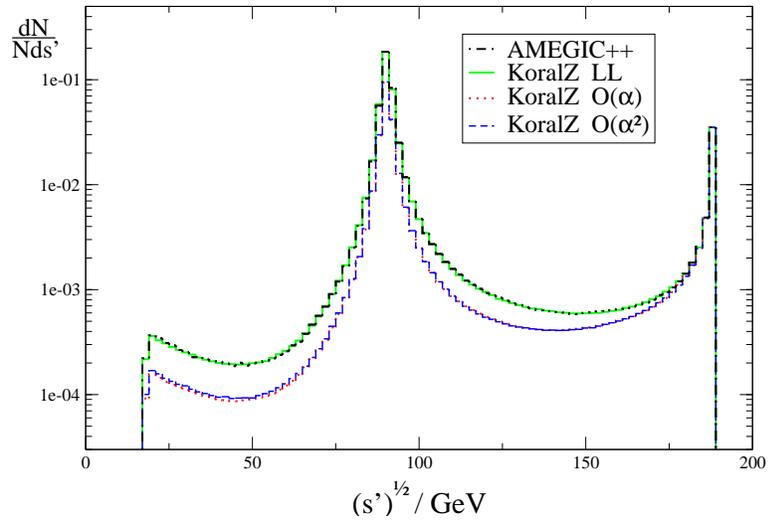}\\[1cm]
\includegraphics[width=7cm,angle=270]{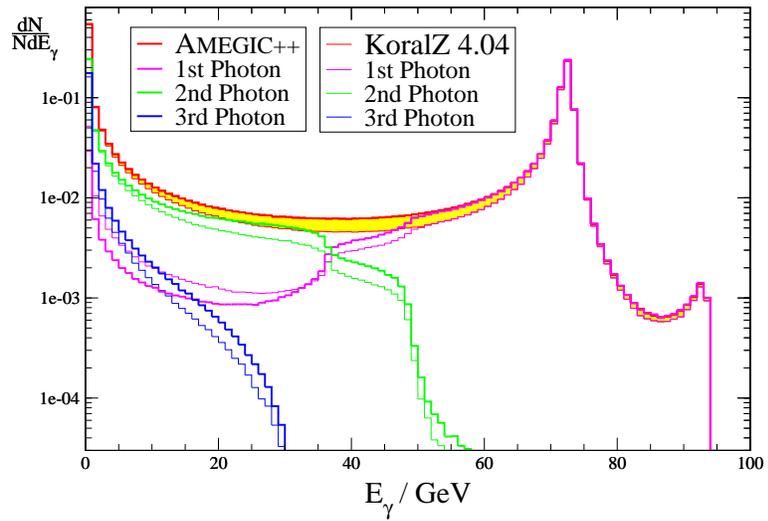}
\caption{\AME and {\tt KoralZ} are compared in the $s'$-distribution
at 189 GeV CMS-energy on the upper panel. On the lower panel the
corresponding energy distribution of the ISR-photons is
displayed.\label{ISR}}  
\end{center}
\end{figure}

At the beginning of every event generation the initial state has to be 
defined. At present our package supports $e^+e^-$ as colliding particles,
but due to initial state radiation of photons the energy as well as
the momentum are not fixed yet. Different approaches describing
the subsequent radiation of photons are the structure function ansatz\cite{ISR_struc},
the electromagnetic shower\cite{Pythia} and the Yennie-Frautschi-Suura (YFS)
scheme\cite{YFS}. Within \AME the first and the last version are
implemented. The structure function ansatz considers the effect of
diminishing the electron energies by initial state photons without 
generating them explicitly. However, the YFS-approach allows a
direct generation of photons in a theoretical well defined way up to
an arbitrary order of $\alpha_{\rm QED}$. In the present state we have
implemented this scheme in the soft photon limit, i.e. an
exponantiation of all effects to leading logarithmic order. In
Fig.\,\ref{ISR} one can see, that \AME agrees with {\tt
KoralZ}\cite{KoralZ} to the considered order.  Further more we
display the effect of higher order corrections, which lead to
significant changes. Hence, we will extend our treatment of
initial state radiation accordingly.                     

\section{Matrix elements}

\begin{figure}[t]
\begin{center}
\includegraphics[width=7cm,angle=270]{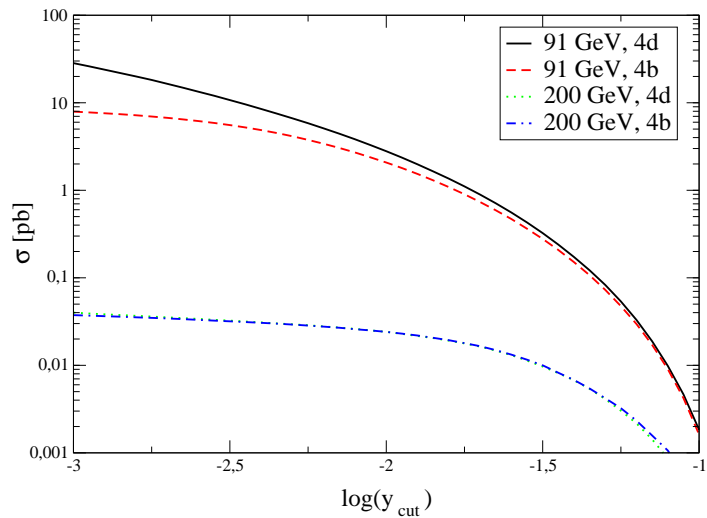}
\caption{The 4 jet rate for four massless and massive quarks at 91 GeV and
200 GeV CMS-energy are displayed.\label{ME}} 
\end{center}
\end{figure}

Now, the inital state is set and the event generation proceeds
with the determination of the jet structure. Jets are defined within
different schemes, e.g. the DURHAM\cite{Durham} and the JADE\cite{Jade}
cluster algorithm. Utilizing these schemes the matrix elements are
regularized, i.e. the soft and collinear divergencies are
avoided. Within \APA different matrix element generators are
supplied, namely {\tt Excalibur}, {\tt Debrecen}\cite{Deb} and
\AME. They all differ in their field of application.  Our
prefered choice is \AME, which is applicable for all standard model
tree level processes up to 6 massive outgoing particles. Its fully
automatic procedure for the determination and integration of the
amplitudes can be divided into three major steps:
\begin{enumerate}
\item The Feynman diagrams are achieved through the mapping of the 
eligible vertices onto tree topologies. 
\item The diagrams are translated into helicity amplitudes and stored
into word-strings for easy evaluation. 
\item Integrating the amplitudes with Rambo\cite{RAMBO},
Sarge\cite{SARGE} or a multi-channel 
\clearpage
approach\cite{MCHAN} (which can
include the former ones) the total cross section is derived. 
\end{enumerate}  
For further details we refer the reader to a more concise description
of our program\cite{AME}. A comparison
between massless and massive four jet cross sections as evaluated with
\AME is depicted in Fig.\,\ref{ME}.
 
\section{Combining matrix elements and parton shower}

Once the jet structure is established by the matrix elements a parton
shower should evolve these different jets. Since particles calculated
via a hard matrix element are naturally on their mass-shell and a
parton shower can handle off-shell particles only, it is obvious that a
scheme for combining these two steps is indispensable. Moreover such a
procedure should take advantage of the virtues of matrix elements,
i.e. the description of jet correlations, and parton shower,
i.e. the evolution of jets. This can be achieved following four steps:
\begin{figure}[t]
\begin{center}
\includegraphics[width=7cm,angle=270]{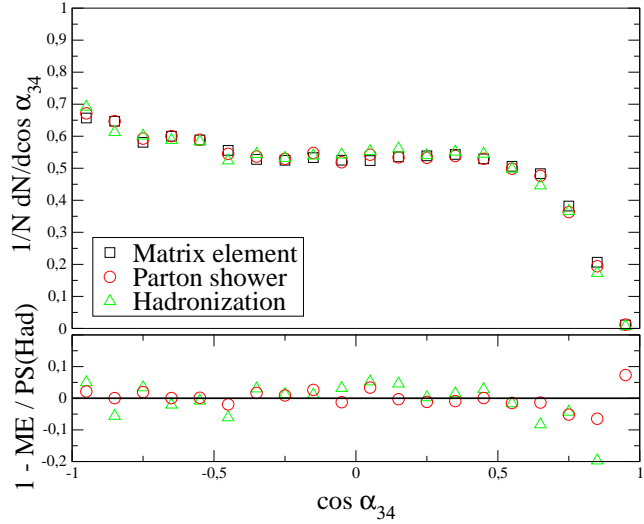}
\caption{\label{4jettop} The $\alpha_{34}$ (the angle between the
lowest energy jets) angle for QCD four jet events at $\sqrt{s}={\rm 206\,GeV}$.} 
\end{center}
\end{figure}
\begin{enumerate}
\item The number and the flavour of the outgoing jets are determined
utilizing the different matrix elements. 
\item The kinematical configuration is chosen according to the matrix
element. An extra weight appears, when higher order corrections are
taken into account, i.e. a combination of rescaled coupling constants
$\alpha_S(y_{\rm cut} s)$ and Sudakov form factors care for an exact
treatment of leading logarithms\cite{CKKW}. This guaranties a smooth
transition of the kinematics from the matrix element to the parton
shower regime. 
\item One of the contributing Feynman diagrams is chosen in order to
gain the colour configuration of the event. The probability for the
selections can be obtained for instance in a parton shower like
manner\cite{JPG,APP}.        
\item A history of the parton branchings is deduced from
the chosen Feynman diagram. Now the partons can be provided with
virtual mass utilizing the Sudakov form factor originating from the
parton shower. 
\end{enumerate}

The success of this combination scheme is especially reflected in four
jet events, see Fig.\,\ref{4jettop}. Needless to say, that this total
agreement between the different phases of event generation could not
be achieved using a parton shower starting from two partons only. A
detailed comparison of the four jet angles between the different event
generators is presented in\cite{yr}.
 
\section{Parton shower and fragmentation}

After all partons gained a virtual mass the evolution of jets can
proceed. Different schemes according to different approximations are
implemented in \APA, i.e. the ordering by virtualities (LLA) and
angles (MLLA). Further details can be found in many textbooks, see for
instance\cite{ESW96}. In addition, azimuthal correlations between the
different planes of parton branchings are taken into account.

Subsequently the outgoing partons have to be hadronized. This is
performed with the help of the Lund-string model\cite{} provided by
Pythia\cite{Pythia}.  

\begin{figure}[t]
\begin{center}
\includegraphics[width=10cm]{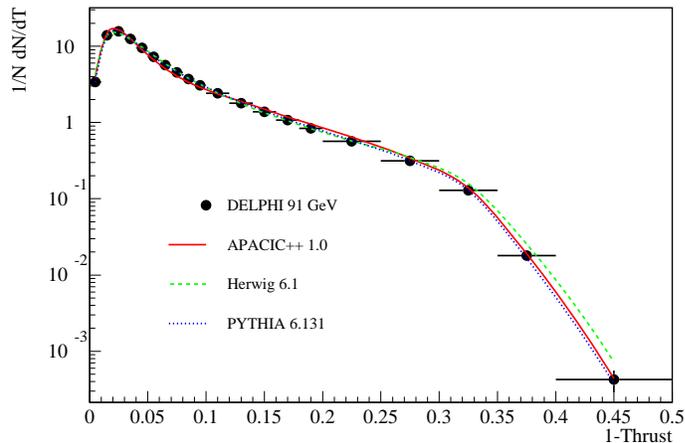}
\caption{The eventshape variable thrust at
$\sqrt{s}={\rm 91\, GeV}$.\label{eventshape}} 
\end{center}
\end{figure}
   
\section{Results}

\begin{figure}[t]
\begin{center}
\includegraphics[width=10cm]{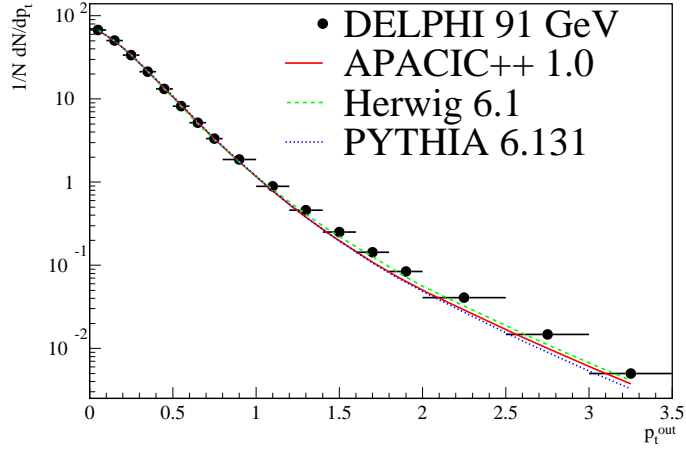}
\caption{The transversal momentum $p_\perp^{\rm out}$ 
at $\sqrt{s}={\rm 91\, GeV}$.\label{momenta}}
\end{center}
\end{figure}

\begin{figure}[h]
\begin{center}
\includegraphics[width=10cm]{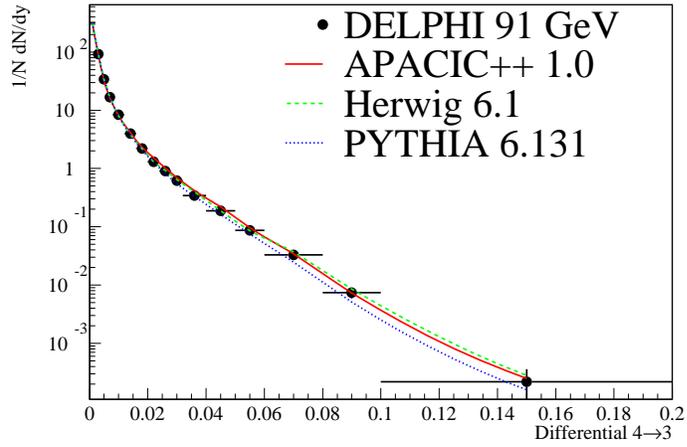}\\
\includegraphics[width=10cm]{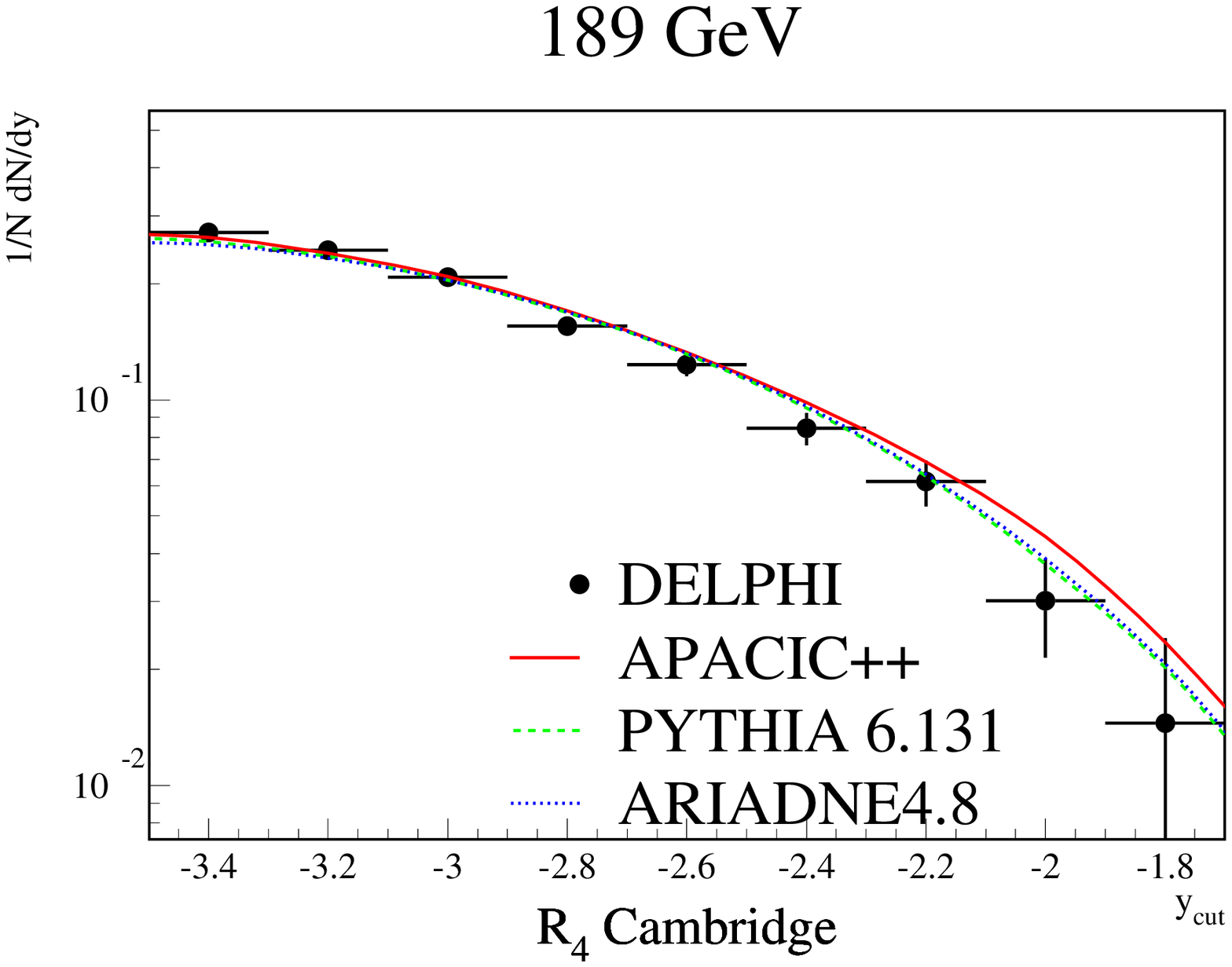}\\
\caption{The energy extrapolation from the differential three jet
rate from 91 GeV (upper panel) to 189 GeV (lower panel).\label{enext}}
\end{center}
\end{figure}

We performed a comparison between Ariadne\cite{Ariadne}, Herwig\cite{Herwig},
Pythia, our event generator \AA and the data of the
DELPHI collaboration at different CMS-energies. In
Fig.\,\ref{eventshape} the thrust distribution (an event shape
variable) at 91 GeV is displayed 
and shows a good overall agreement with the data. The transversal
momentum $p_\perp^{\rm out}$ could not be
described correctly in the high momentum region. This is seemingly a
common feature of all event generators, see Fig.\,\ref{momenta}. 
Even though our program includes the full information of matrix
elements an energy extrapolation has been achieved, see
Fig.\,\ref{enext}. This is an important feature of \AA, since a pure
matrix element generator does not have this property. 

We conclude, that we reached the aim of providing an event generator,
which is able to describe multijet topologies with a proper energy
scaling.  
 
\section*{Acknowledgments}

F.K., R.K. and A.S.  would like to thank J. Drees, K. Hamacher and
U. Flagmeyer for helpful discussions. During the process of tuning the
{\tt APACIC++}--parameters to experimental data by U. Flagmeyer we
were able to identify and cure some shortcomings and bugs of the program.

For R.K. and F.K. it is a pleasure to thank L. Lonnblad and
T. Sjostrand for valuable comments and S. Catani and B. Webber for the
pleasant collaboration on the combination of matrix elements and
parton showers. 

This work is supported by BMBF.


\clearpage
\begin{thebibliography}{99}
\bibitem{APA} R. Kuhn, F. Krauss, B. Ivanyi, G. Soff; hep-ph/0004270, 
	      accepted by Comp. Phys. Commun..
\bibitem{AME} F. Krauss, R. Kuhn, G. Soff in preparation.
\bibitem{ISR_struc} F. A. Berends, R. Pittau, R. Kleiss, Nucl. Phys. B426 (1994) 344.
\bibitem{Pythia} T. Sjostrand, Comp. Phys. Commun. 82 (1994) 74.
\bibitem{YFS} D. R. Yennie, S. C. Frautschi and H. Suura, Ann. Phys. 13 (1961) 379.
\bibitem{Exc} F. A. Berends, R. Pittau, R. Kleiss, Comp. Phys. Commun. 85 (1995) 437.
\bibitem{KoralZ} S. Jadach, B.F.L. Ward, Z. Was,
Comp. Phys. Commun. 124 (2000) 233.
\bibitem{Durham} S. Catani, Yu. L. Dokshitzer, M. Olsson, G. Turnock, 
                 B. R. Webber, Phys. Lett. B269 (1991) 432.
\bibitem{Jade} Jade--Collaboration, S. Bethke et al.,
               Phys. Lett. B213 (1988) 235.
\bibitem{Deb} Z. Nagy, Z. Trocsanyi, Nucl. Phys. B, Proc. Suppl. 64 (1998) 63.
\bibitem{RAMBO} R. Kleiss, W. J. Stirling, S. D. Ellis,
                   Comp. Phys. Commun. 40 (1986) 359;
                R. Kleiss, W. J. Stirling, Nucl. Phys. B385 (1992) 413.
\bibitem{SARGE} P.~D.~Draggiotis, A.~van Hameren and R.~Kleiss,
	        Phys. Lett. B483 (2000) 124. 
\bibitem{MCHAN} R. Kleiss, R. Pittau, Comp. Phys. Commun. 83 (1994) 141.
\bibitem{CKKW} S.~Catani, F.~Krauss, R.~Kuhn, B.~R.~Webber, in preparation
\bibitem{JPG}  F. Krauss, R. Kuhn, G. Soff, J. Phys. G 26 (2000) L11.
\bibitem{APP}  F. Krauss, R. Kuhn, G. Soff, Acta Phys. Pol. B30 (1999) 3875.
\bibitem{yr}     A. Ballestrero et al., hep-ph/0006259, appeared in
	         \textsl{Reports of the working groups on precision
	 	 calculations for LEP2 physics}, CERN 2000-009, p.~137, ISBN 92-9083-171-5
%\bibitem{NR}   O. Nachtmann, A. Reiter, Z. Phys. C16 (1982) 45.
\bibitem{ESW96} R. K. Ellis, W. J. Stirling, B. R. Webber,
                \textsl{QCD and Collider Physics},
                Cambridge Monographs on Particle Physics, Nuclear Physics
                and Cosmology, Cambridge University Press, 1. Edition (1996).
\bibitem{Ariadne} L. Lonnblad, Comp. Phys. Commun. 71 (1992) 15.
\bibitem{Herwig} G. Marchesini, B. R. Webber, G. Abbiendi, I. G. Knowles,
                 M. H. Seymour, L. Stanco, Comp. Phys. Commun. 67 (1992)
                 465.                                                             
\end{thebibliography}
\end{document}